\documentclass[letterpaper]{article} 
\usepackage{aaai25}  
\usepackage{times}  
\usepackage{helvet}  
\usepackage{courier}  
\usepackage[hyphens]{url}  
\usepackage{graphicx} 
\usepackage{natbib}  
\usepackage{caption} 
\usepackage{algorithm}
\usepackage{algorithmic}

\usepackage{amsmath}
\usepackage{amssymb}
\usepackage{subcaption}

\usepackage{newfloat}
\usepackage{listings}
\usepackage{booktabs,makecell, multirow, tabularx}
\usepackage{newfloat}
\usepackage{listings}

\newtheorem{theorem}{Theorem}
\newtheorem{lemma}[theorem]{Lemma}
\newtheorem{corollary}[theorem]{Corollary}

\DeclareCaptionStyle{ruled}{labelfont=normalfont,labelsep=colon,strut=off} 
\floatstyle{ruled}
\newfloat{listing}{tb}{lst}{}
\floatname{listing}{Listing}
\nocopyright 

\title{FedFQ: Federated Learning with Fine-Grained Quantization}
\author{
    Haowei Li,
    Weiying Xie,
    Hangyu Ye,
    Jitao Ma,
    Shuran Ma,
    Yunsong Li
}
\affiliations{
Xidian University\\

    No. 2 South Taibai Road, Xi’an, Shaanxi, China\\
    hw\_li@stu.xidian.edu.cn
}

\usepackage{bibentry}

\begin{document}

\maketitle

\begin{abstract}
	Federated learning (FL) is a decentralized approach, enabling multiple participants to collaboratively train a model while ensuring the protection of data privacy. The transmission of updates from numerous edge clusters to the server creates a significant communication bottleneck in FL. Quantization is an effective compression technology, showcasing immense potential in addressing this bottleneck problem. The Non-IID nature of FL renders it sensitive to quantization. Existing quantized FL frameworks inadequately balance high compression ratios and superior convergence performance by roughly employing a uniform quantization bit-width on the client-side. In this work, we propose a communication-efficient FL algorithm with a fine-grained adaptive quantization strategy (FedFQ). FedFQ addresses the trade-off between achieving high communication compression ratios and maintaining superior convergence performance by introducing parameter-level quantization. Specifically, we have designed a Constraint-Guided Simulated Annealing algorithm to determine specific quantization schemes. We derive the convergence of FedFQ, demonstrating its superior convergence performance compared to existing quantized FL algorithms. We conducted extensive experiments on multiple benchmarks and demonstrated that, while maintaining lossless performance, FedFQ achieves a compression ratio of 27$\times$ to 63$\times$ compared to the baseline experiment.
\end{abstract}

%

\section{Introduction}

Federated Learning (FL) is a decentralized machine learning approach that enables multiple participants to collaboratively train a global model while preserving data localization for privacy reasons \cite{mcmahan2017communication}. In FL, a large number of clients need to upload their local models to the server. Due to the unstable nature and limited bandwidth of edge devices serving as FL clients, the issue of communication bottlenecks poses a significant challenge that can severely impact or even undermine the effectiveness of FL \cite{chen2021communication,kairouz2021advances,xiong2023feddm}. \begin{figure}[htpb]
	\centering
	\includegraphics[width=1.05\columnwidth]{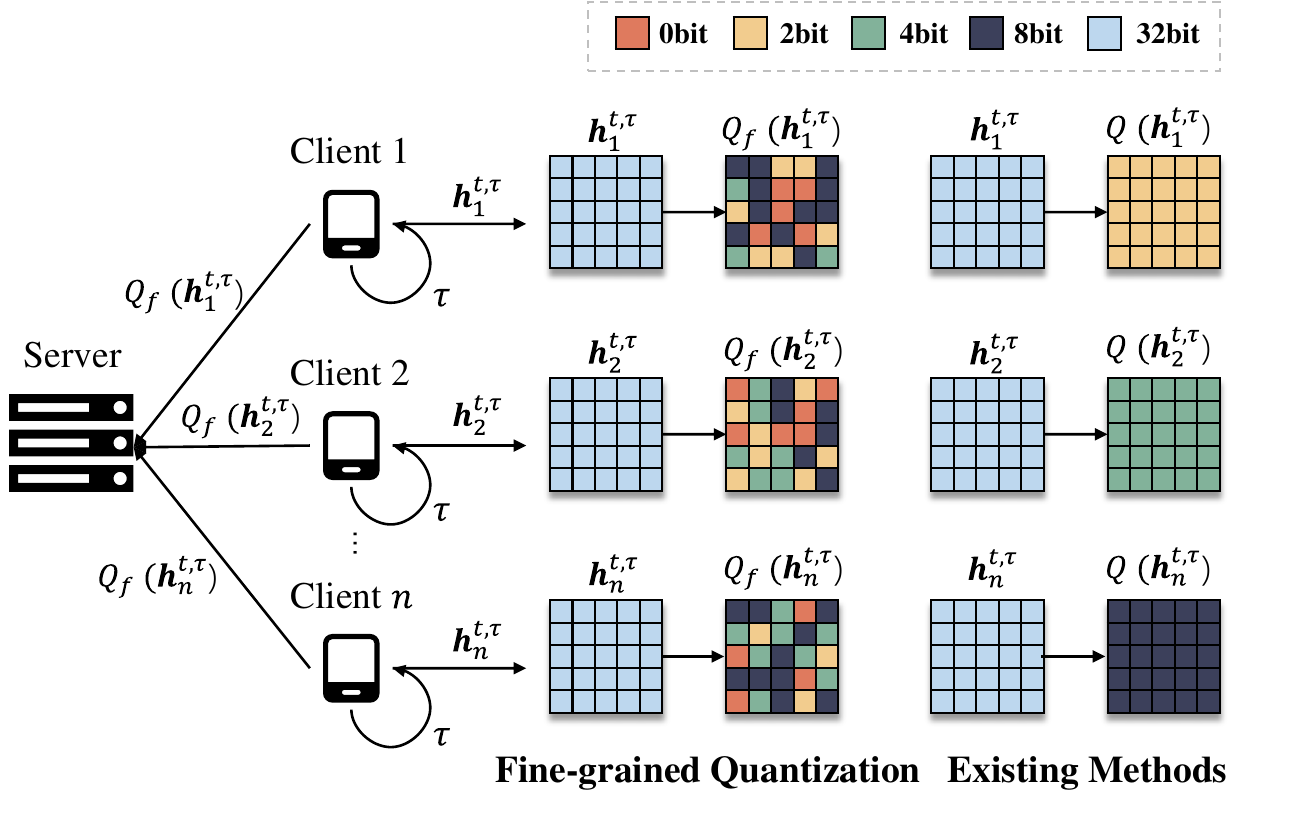}
	\caption{The schematic diagram of FedFQ. $\boldsymbol{h}$ represents local model updates. FedFQ's fine-grained quantization is an adaptive quantization technique targeting individual parameters in the parameter space, as opposed to the existing methods of uniformly quantizing the entire parameter space.}
	\label{p0}
	\vspace{-5pt}
\end{figure}

Addressing this problem has become a focal point of research in recent years.  Quantization, as an effective compression technique, can be utilized to reduce the amount of data uploaded from clients to servers in FL. Early proposed quantization methods \cite{konevcny2016federated,reisizadeh2020fedpaq} uniformly quantize all updates to a fixed bit-width and recent studies have explored allocating different quantization bit-widths for each clients but using a single bit-width for local updates within the same client \cite{mao2022communication}. However, the existing quantized FL algorithm cannot simultaneously achieve high communication compression ratios and superior convergence performance. In FL, the data exhibits Non-IID characteristics. This introduces ``model drift'' which hampers convergence. Unfortunately, the utilization of quantization operators further exacerbates ``model drift'' by introducing additional information loss \cite{sattler2019robust,yang2021cfedavg}. Moreover, quantization in FL often integrates with widely used local methods such as FedAvg, where the transmitted data is a result of multiple local updates. Therefore, the data has lower redundancy \cite{reisizadeh2020fedpaq}. The aforementioned factors render FL highly sensitive to quantization with high communication compression ratios.  It requires quantization algorithms to have minimal information loss. Existing methods crudely quantize the client's local updates to a single bit-width, disregarding the parameter distribution information \cite{huang2021rethinking} within the clients.

In this paper, we propose a communication-efficient FL algorithm with a fine-grained adaptive quantization strategy, called FedFQ. The core idea is to enable customized bit-width selection at the individual parameter level, as shown in Figure \ref{p0}. FedFQ leverages the specific information of the parameter distribution within clients, ensuring maximum adaptivity in quantization. We theoretically derive the convergence of FedFQ in both strongly convex and non-convex settings, and demonstrate its superior convergence performance compared to existing quantized FL algorithms.
Furthermore, we devise a Constraint-Guided Simulated Annealing algorithm with polynomial time complexity to determine the optimal fine-grained quantization strategy. Our FedFQ algorithm follows the design of FedAvg \cite{mcmahan2017communication}, where in each communication round, only a subset of clients uploads the results of multiple model updates to achieve more efficient communication. Through theoretical analysis and extensive experiments on benchmarks, including image classification tasks on CIFAR10 \cite{krizhevsky2009learning} with SimpleCNN \cite{mcmahan2017communication} and VGG11 \cite{simonyan2014very}, as well as character prediction tasks on Shakespeare with NextCharLSTM \cite{kim2016character}, we demonstrate the superior performance of FedFQ in FL scenarios. Compared to existing communication-efficient FL algorithms, FedFQ achieves the lowest communication overhead while maintaining performance without any loss. Our contributions can be summarized as follows:

\begin{itemize}
	\item We propose FedFQ, a communication-efficient FL algorithm with a fine-grained adaptive quantization strategy. To the best of our knowledge, our quantization strategy is the first to allow for customized selection of bit-width at the individual parameter level.
	
	\item We theoretically derive the superior convergence performance of FedFQ. Furthermore, we devise a Constraint-Guided Simulated Annealing algorithm with polynomial time complexity to determine the optimal fine-grained quantization strategy.
	
	\item We empirically evaluate FedFQ through extensive experiments. FedFQ addresses the trade-off between high communication compression ratios and superior convergence performance. With the premise of maintaining lossless performance, FedFQ achieves a compression ratio of 27$\times$ to 63$\times$ compared to the baseline experiments.
\end{itemize}

\section{Related Work}

\textbf{Strategies for Communication} Several studies have effectively reduced communication overhead in FL. Firstly, some algorithms aim to minimize the frequency of model updates. FedAvg \cite{mcmahan2017communication} allowed clients to perform multiple local paramer updates before each communication round. FedProx \cite{li2020federated} enhanced FedAvg's performance on Non-IID data.
Secondly, the communication volume could be reduced by selecting a subset of clients instead of involving all clients. FedAvg \cite{mcmahan2017communication} randomly selected a subset of clients for each communication round. FedCS \cite{nishio2019client} and E3CS \cite{huang2022stochastic} selected clients with the highest model iteration efficiency for aggregated updates. 
Moreover, compression methods such as sparsification and quantization can be used to reduce the uploaded data's size.

\textbf{Sparsification and Quantization} Sparsification selects specific components of local updates for transmission. Strom \textit{et al.} \cite{strom2015scalable} and Aji \textit{et al}. \cite{aji2017sparse} proposed transmitting only gradients above a threshold. DGC \cite{lin2017deep} improved upon Strom \textit{et al.}'s approach by adding momentum correction and small gradient accumulation, achieving better performance. Song \textit{et al.} \cite{song2021communication} not only focused on large-magnitude gradients but also increased the sampling rate for smaller gradients. 
Quantization aims to transmit model updates with reduced precision. SignSGD \cite{bernstein2018signsgd} compressed gradients to 1-bit. QSGD \cite{alistarh2017qsgd} introduced random uniform quantization with varying bit-widths. AQSGD \cite{faghri2020adaptive} dynamically adjusted the quantization scheme in real-time to speed up training. Some studies combine quantization with sparsification to benefit from both methods \cite{yan2022ac,basu2019qsparse}.

\textbf{Compression in FL} Sparsification has been applied in FL algorithms. Sattler \textit{et al.} \cite{sattler2019robust} proposed a sparse ternary compression method STC that provides compression for both the uplink and downlink communication in FL. Hu \textit{et al.} \cite{ijcai2021p202} combined random sparsification with gradient perturbation at each client. Bibikar \textit{et al.} \cite{bibikar2022federated} introduced dynamic sparsification in FL training to adapt to the Non-IID nature of FL more flexibly. Regarding quantization methods in FL, which is also the focus of this paper, Konecny \textit{et al.} \cite{konevcny2016federated} combined quantization with random rotations and sub-sampling. Reisizadeh \textit{et al.} \cite{reisizadeh2020fedpaq} introduced random uniform quantization based on FedAvg. Shlezinger \textit{et al.} \cite{shlezinger2020uveqfed} proposed the concept of vector quantization. Mao \textit{et al.} \cite{mao2022communication} adaptively selected the quantization bit-width at each client. Existing quantization methods often use a single bit-width for local updates at the client, ignoring the parameter distribution information of local updates. Compared to previous work, FedFQ is the first to allow for customized selection of bit-width at the individual parameter level, ensuring optimal convergence performance at high compression ratios.

\section{Fine-grained And Adaptive Quantization Strategies}
In this section, we begin by presenting the formulation for FL. We then provide an unbiasedness and bounded variance analysis of the fine-grained quantization in FedFQ, and then analyze the convergence error upper bound of FedFQ. We demonstrate that our algorithm exhibits superior convergence performance compared to conventional quantization algorithms. Finally, we elaborate on the specific algorithm of fine-graine quantization. For a detailed explanation of the content in this section, please refer to the appendix.
\subsection{FedAvg Formulation}
A typical FL system consists of $n$ clients and a central server. Each client has a local dataset with
$m$ data samples, and all clients collaboratively train a global
model $\boldsymbol{\theta}$ on the collection of their local datasets under the orchestration of the central server. The clients in FL aim to find the optimal global model $\boldsymbol{\theta}$ by solving the following empirical risk minimization problem while keeping their data locally:
\begin{equation}
	\min _{\boldsymbol{\theta} \in \mathbb{R}^d} f(\boldsymbol{\theta}):=\frac{1}{n} \sum_{i=1}^n f_i(\boldsymbol{\theta}),
\end{equation}
where $f_i(\boldsymbol{\theta})$ represents the loss function of
$i$-th client.

FedAvg is a communication-efficient FL algorithm that involves parallel execution of multiple local updates on selected clients, followed by aggregation of these client model updates. Specifically, FedAvg consists of $K$ communication rounds, and the execution process in each round is as follows:
At the beginning of round $t \in\{0, \ldots, K-1\}$, the server selects a subset of clients $\mathcal{W} \subseteq[n]$ to participate and sends them the latest global model $\boldsymbol{\theta}_t$. Subsequently, each client $i \in \mathcal{W}$ initializes a local model $\boldsymbol{\theta}_i^{t, 0}$ and performs $\tau$ local updates, as illustrated below:
\begin{equation}
	\boldsymbol{\theta}_i^{t, s+1}=\boldsymbol{\theta}_i^{t, s}-\eta_{t,s} \boldsymbol{g}_i^{t, s}, \quad s=0,1, \ldots, \tau-1
\end{equation}
where $\eta_{t,s}$ is the local learning rate and  $\boldsymbol{g}_i^{t, s}$ represents the stochastic gradient. Then, each client uploads its final local model update $\boldsymbol{\theta}_i^{t, \tau}-\boldsymbol{\theta}_t$ to the server. The server aggregates the local updates from all participating clients and updates the global model as follows:
\begin{equation}
	\boldsymbol{\theta}_{t+1}=\boldsymbol{\theta}_t+\frac{1}{|\mathcal{W}|} \sum_{i \in \mathcal{W}}\left(\boldsymbol{\theta}_i^{t, \tau}-\boldsymbol{\theta}_t\right).
\end{equation}

\subsection{The Convergence of FedFQ}

Our FedFQ introduces parameter-level mixed-precision quantization to FedAvg. After selecting clients to local update multiple times, we apply a quantization strategy to the updates $\boldsymbol{h}_i^{ t, \tau}=\boldsymbol{\theta}_i^{t, \tau}-\boldsymbol{\theta}_t$. The server then aggregates the quantized model updates as follows:
\begin{equation}
	\boldsymbol{\theta}_{t+1}=\boldsymbol{\theta}_t+\frac{1}{|\mathcal{W}|} \sum_{i \in \mathcal{W}} Q_f\left(\boldsymbol{h}_i^{ t, \tau}\right),
\end{equation}
where $Q_f\left(\cdot\right)$ represents our mixed-precision quantization strategy. 

We firstly discuss the general form of random uniform quantization \cite{alistarh2017qsgd}, which can be formulated as follows:
\begin{equation}
	Q\left(\boldsymbol{h}_i^{ t, \tau}\right)=\left\|\boldsymbol{h}_i^{ t, \tau}\right\|_2 \cdot \operatorname{sign}\left(\boldsymbol{h}_i^{ t, \tau}\right) \cdot \xi_i\left(\boldsymbol{h}_i^{ t, \tau}, s\right),
\end{equation}
$\xi_i\left(\boldsymbol{h}_i^{ t, \tau}, s\right)$ is an unbiased stochastic function that maps scalar $\frac{\left|\left(\boldsymbol{h}_i^{ t, \tau}\right)_j\right|}{\left\|\boldsymbol{h}_i^{ t, \tau}\right\|_2}$ to one of the values in set $\{0,1 / s, 2 / s, \ldots, s / s\}$, $j$ represents the $j$-th element in $\boldsymbol{h}_i^{ t, \tau}$, while $s = 2^{b-1}$ denotes the quantization level. $b$ represents the quantization bit-width.
For random uniform quantization, we extend the theory from \cite{alistarh2017qsgd} to obtain the following lemma:

\begin{lemma} 
	(Unbiasness and Bounded Variance of Random Uniform Quantization): The random quantizer $Q(\cdot)$ is unbiased and its variance grows with the squared of l2-norm of its argument, i.e.,
	\begin{align}
		\mathbb{E}[Q(\boldsymbol{h}_i^{ t, \tau})] &= \boldsymbol{h}_i^{ t, \tau} \label{6},\\
		\mathbb{E}\left[\|Q(\boldsymbol{h}_i^{ t, \tau})-\boldsymbol{h}_i^{ t, \tau}\|_2^2 \right] &\leq q\|\boldsymbol{h}_i^{ t, \tau}\|_2^2 \label{7}, \quad q=\frac{d}{4^b}.
	\end{align}
\end{lemma}
Here, $d$ represents the length of $\boldsymbol{h}_i^{ t, \tau}$. A higher quantization bit-width leads to a smaller upper bound of the variance. 

FedFQ adopts a more fine-grained quantization, selecting the most appropriate bit width for each individual element.
We quantize the $j$-th element of $\boldsymbol{h}_i^{ t, \tau}$ using $b_j$ bits. For the purpose of convenience in description, we pad each element to the original data length $d$. Therefore, we have $\boldsymbol{h}_i^{t, \tau}=\sum_{j=1}^d\left(\boldsymbol{h}_i^{t, \tau}\right)_j$ and $\left\|\boldsymbol{h}_i^{t, \tau}\right\|_2^2=\sum_{j=1}^d\left|\left(\boldsymbol{h}_i^{t,\tau}\right)_j\right|^2$.
We denote the quantization strategy of FedFQ as $Q_f(\cdot)$, and its expected value  satisfy the following equation:
\begin{equation}
	\resizebox{.71\linewidth}{!}{$
		\mathbb{E}\left[Q_f\left(\boldsymbol{h}_i^{t, \tau}\right)\right]=\mathbb{E}\left[\sum_{j=1}^d Q\left(\left(\boldsymbol{h}_i^{t, \tau}\right)_j\right)\right],
		$}\label{8}
\end{equation}
and its variance satisfies:
\begin{equation}
	\resizebox{.91\linewidth}{!}{$
		\displaystyle
		\mathbb{E}\left[\left\|Q_f\left(\boldsymbol{h}_i^{t, \tau}\right)-\boldsymbol{h}_i^{t, \tau}\right\|_2^2\right]=\mathbb{E}\left[\sum_{j=1}^d\left|Q\left(\left(\boldsymbol{h}_i^{t, \tau}\right)_j\right)-\left(\boldsymbol{h}_i^{t, \tau}\right)_j\right|^2\right]
		$}.\label{9}
\end{equation}

According to Eq. (\ref{6}) and Eq. (\ref{7}), we can obtain the following inequality: \\
$\mathbb{E}[Q(\left(\boldsymbol{h}_i^{t, \tau}\right)_j)] = \left(\boldsymbol{h}_i^{t, \tau}\right)_j, \mathbb{E}\left[\left|Q\left(\left(\boldsymbol{h}_i^{t, \tau}\right)_j\right)-\left(\boldsymbol{h}_i^{t, \tau}\right)_j\right|^2 \right] \leq \frac{d}{4^{b_j}}\left|\left(\boldsymbol{h}_i^{t,\tau}\right)_j\right|^2$.
By substituting them into Eq. (\ref{8}) and Eq. (\ref{9}), we can obtain the following expressions:
\begin{theorem} 
	(Unbiasness and Bounded Variance of FedFQ): The quantizer of FedFQ $Q_f(\cdot)$ is unbiased and its variance grows with the squared of l2-norm
	of its argument, i.e.,
	\begin{align}
		\mathbb{E}[Q_f(\boldsymbol{h}_i^{ t, \tau})] &= \boldsymbol{h}_i^{ t, \tau} \label{10},\\
		\mathbb{E}\left[\|Q_f(\boldsymbol{h}_i^{ t, \tau})-\boldsymbol{h}_i^{ t, \tau}\|^2 \right] &\leq q_f\|\boldsymbol{h}_i^{ t, \tau}\|_2^2, \\
		q_f = \sum_{j=1}^d \frac{d}{4^{b_j}}& \frac{\left|\left(\boldsymbol{h}_i^{t, \tau}\right)_j\right|^2}{\left\|\boldsymbol{h}_i^{t, \tau}\right\|_2^2}.
		\label{11}
	\end{align}
\end{theorem}
By comparing Eq. (\ref{11}) and Eq. (\ref{7}), we observe that the conventional quantization method is a specific instance of FedFQ with a singular quantization bit-width. $\min q_f \leq \min q.$ 

To better explore the mathematical relationship between $q_f$ and $q$, we consider the following scenario: When the communication budget, \textit{i.e.}, the total number of bits, is given, we assume that there are $d_1$ elements with increased bit-width, with increments $x_1, x_2, \ldots, x_{d_1}$, and $d_2$ elements with decreased bit-width, with decrements $y_1, y_2, \ldots, y_{d_2}$, satisfying $x_1+x_2+\ldots+x_{d_1}=y_1+y_2+\ldots+y_{d_2}$ and $d=d_1+d_2$. The upper bound of the variance in this case is:
\begin{equation} 
	q_f = \sum_{j=1}^{d_1}\frac{d}{4^{b+x_j}} \frac{\left|\left(\boldsymbol{h}_i^{t,\tau}\right)_j\right|^2}{\left\|\boldsymbol{h}_i^{t,\tau}\right\|_2^2}+\sum_{k=1}^{d_2}\frac{d}{4^{b-y_k}} \frac{\left|\left(\boldsymbol{h}_i^{t,\tau}\right)_j\right|^2}{\left\|\boldsymbol{h}_i^{t,\tau}\right\|_2^2}.
\end{equation} 
When the following inequality is satisfied, $q_f < q$:
\begin{equation}
	\sum_{j=1}^{d_1} \frac{4^{x_j}-1}{4^{x_j}}\left|\left(\boldsymbol{h}_i^{t,\tau}\right)_j\right|^2>\sum_{k=1}^{d_2}\left(4^{y_k}-1\right)\left|\left(\boldsymbol{h}_i^{t,\tau}\right)_k\right|^2.
\end{equation}
This reveals the following corollary:
\begin{corollary}
	(1) The greater the discrepancy in parameter magnitudes within the parameter space $\boldsymbol{h}_i^{ t, \tau}$, the higher the performance of FedFQ. (2) Only sufficiently large components are worth allocating higher bit widths.
	\label{corollary}
\end{corollary}
During training, model parameters often exhibit diverse distributions \cite{huang2021rethinking}. Moreover, due to the Non-IID nature of data across clients, data distributions vary significantly between them. Therefore, FedFQ can significantly reduce the upper bound of quantization variance. In the limit, it can be shown that $q_f \leq \frac{1}{4^k} q$, where $k$ is related to the total available bit widths.

In fact, quantization's effect on federated learning convergence closely ties to the quantizer's variance upper bound. We explore this concerning both strongly convex smooth and non-convex smooth loss functions, drawing the following conclusions:
\begin{theorem}
	For strongly convex smooth functions,the expected error: 
	\begin{equation}
		\begin{aligned}
			\mathbb{E}\left\| \boldsymbol{\theta}_t - \boldsymbol{\theta}^* \right\|^2 &\leq k_0 \frac{\tau^2}{T^2} + \left(k_1 q + c_1\right) \frac{\tau}{T} \\
			& \quad + k_2 \frac{(\tau-1)^2}{T} + \left(k_3 q + c_3\right) \frac{\tau-1}{T^2}.
		\end{aligned}
	\end{equation}
	For smooth non-convex loss functions, the following first-order stationary condition holds: 
	\begin{equation}
		\begin{aligned}
			\frac{1}{T} \sum_{k=0}^{T-1} \sum_{t=0}^{\tau-1} \mathbb{E}\left\|\nabla f\left(\overline{\boldsymbol{\theta}}^{t, s}\right)\right\|^2 \leq & \frac{k_0}{\sqrt{T}}+\left(k_1 q+C_1\right) \frac{1}{\sqrt{T}} \\
			& +k_2 \frac{r-1}{T}.
		\end{aligned}
	\end{equation}
	Here, $\tau$ represents the local update count, while $T$ denotes the total number of iterations with $T=K*\tau$. The parameters $k$ and $c$ are associated with the step size, the total number of clients, and the participating clients.  $k_0$ is related to the initial error $\mathbb{E}\left\|\boldsymbol{\theta}_{t_0}-\boldsymbol{\theta}^*\right\|^2$ for convex function and $f\left(\boldsymbol{\theta}_0\right)-f^*$ for non-convex function.
	\label{theorem4}
\end{theorem}

Here, $k$ and $c$ are not constants but are linked to other training parameters in federated learning. Our focus is solely on how the quantized parameter $q$ affects convergence. Based on Theorem \ref{theorem4}, we can derive an important corollary:
\begin{corollary}
	For an unbiased and bounded quantization operator, the upper bound of variance $q$ is positively correlated with the upper bound of convergence error.
	\label{corollary1}
\end{corollary}

We can consider an extreme case, where $q=0$ and $\tau=1$. For a strongly convex loss function, when $\mathbb{E}\left| \boldsymbol{\theta}_t - \boldsymbol{\theta}^* \right|^2 \leq O\left(\frac{1}{T}\right) + e$ (initial error term), which is close to the convergence speed of Vanilla parallel SGD. However, Increasing $q$ significantly impacts convergence rate, confirming the earlier assertion that quantization operators exacerbate model drift in federated learning. 

\subsection{Quantization Strategy Of FedFQ}
In conclusion, FedFQ possesses the lowest theoretical variance upper bound, thus demonstrating superior convergence performance. Our optimization objective is to minimize the variance upper bound $q_f$. Given a communication budget $B$, the design of FedFQ can be framed as the following optimization problem:
\begin{equation}
	\resizebox{.71\linewidth}{!}{$
		\label{23}
		\begin{aligned}
			& \min \sum_{j=1}^d \frac{d}{4^{b_j}} \frac{\left|\left(\boldsymbol{h}_i^{t, \tau}\right)_j\right|^2}{\left\|\boldsymbol{h}_i^{t, \tau}\right\|_2^2}, \text { s.t. } \sum_{j=1}^d b_j=B.
		\end{aligned}
		$}
\end{equation}

The vast search space resulting from the large parameter size — tens to hundreds of thousands — of client local updates makes direct solutions computationally expensive. Furthermore, the clients in federated learning are often edge devices with limited computational capabilities.

When facing these challenges, conventional search algorithms exhibit limitations. Dynamic programming can provide the global optimal solution but at a high computational cost, while greedy algorithms are computationally efficient but susceptible to local optima. To address these challenges, we propose a \textbf{Constraint-Guided Simulated Annealing (CGSA) algorithm}, as shown in Algorithm \ref{cgsa}.
\begin{algorithm}[tb]
	\caption{Quantization Strategy $Q_f(\cdot)$ using CGSA}
	\label{cgsa}
	\small
	\textbf{Input}: Local updates $\boldsymbol{h}$, budget $B$, bit-widths $b_k \in\{0,2,4,8\}$, initial temperature $T$, cooling rate $\alpha$, min temperature $min\_T$, max iterations $max\_iter$.\\
	\textbf{Output}: Quantized updates $\widetilde{\boldsymbol{h}}$ 
	\begin{algorithmic}[1]
		\STATE Sort indices of $h$ by magnitude: $idx \gets argsort(-\left|h\right|)$
		\STATE $objective \gets \sum_{j=1}^d \frac{d}{4^{b_j}} \left|h_j\right|^2$
		\STATE\textbf{\# Initial solution}
		\FOR{$i = 0$ to $B/2 - 1$}
		\STATE $b[idx[i]] \gets b[idx[i]] + 2$
		\ENDFOR
		\STATE $best\_value \gets objective(b)$
		\WHILE{$T > min\_T$ and $max\_iter > 0$}
		\STATE Clone $b$ to $\text{new\_b}$
		\STATE \textbf{\# Neighborhood search}
		\STATE Randomly select $i<j$
		\IF{$new\_b[idx[i]] > 0$ and $new\_b[idx[j]] < 8$}
		\STATE $new\_b[idx[i]] \gets new\_b[idx[i]] * 2$
		\STATE $new\_b[idx[j]] \gets new\_b[idx[j]] / 2$
		\ENDIF
		\STATE $current\_value \gets objective(new\_b)$
		\STATE $delta = current\_value - best\_value$
		\STATE\textbf{\# Acceptance criterion}
		\IF{$\delta < 0$ or $random(0, 1) < exp(-\delta / T)$}
		\STATE $b \gets new\_b$
		\STATE $best\_value \gets current\_value$
		\ENDIF
		\STATE\textbf{\# Cooling down}
		\STATE $T \gets T*\alpha$, $max\_iter \gets max\_iter-1$
		\ENDWHILE
		\STATE  Quantize using random quantization: $\widetilde{\boldsymbol{h}} \leftarrow Q\left(\boldsymbol{h}, b\right)$
		\STATE \textbf{return} $\widetilde{\boldsymbol{h}}$
	\end{algorithmic}
\end{algorithm}

Simulated Annealing (SA) is a probabilistic algorithm that aims to approximate the global optimum of a given function. It has two main advantages: 1) the ability to probabilistically accept suboptimal solutions to avoid local optima, and 2) the capability to converge to an approximate optimal solution after a finite number of searches from the initial solution. This approach combines computational efficiency with accuracy.

We introduce heuristic initial solutions and directional constraints on neighborhood generation on top of the conventional SA algorithm to further reduce the number of searches. Our basis is Corollary \ref{corollary}. 
\begin{itemize}
	\item \textbf{Initial solution.} We set the available bit-width options to be 2, 4, and 8 bits. Our priority is to assign 2 bits to larger components until reaching the communication upper limit B, with any unallocated components being set to zero, as shown in line 4 of Algorithm \ref{cgsa}.
	\item \textbf{Neighborhood search}
	According to Corollary 3, the convergence direction towards the globally optimal solution is determined: increasing the bit width of larger components and decreasing the bit width of smaller components. Therefore, we constrain the neighborhood search direction to align with this pattern, as shown in line 13 of Algorithm \ref{cgsa}.
\end{itemize}
By imposing this constraint, the search space is significantly reduced, enabling the algorithm to achieve a balance between computational efficiency and accuracy. We have designed an optimal quantization strategy for FedFQ using the CGSA algorithm with polynomial time complexity. Further details regarding the algorithm can be found in the appendix.

\section{Experiments}
The goal of this section is to evaluate the performance of FedFQ on several  benchmarks. We conducted ablation experiments and comparative experiments. The ablation experiments involved comparing the performance of single-precision quantization and mixed-precision quantization, while the comparative experiments focused on comparing the performance of FedFQ with existing quantized FL algorithms. For each task, we provide performance comparisons separately for both IID and Non-IID scenarios. In addition, we also conducted performance evaluations of FedFQ in realistic network environments.

\subsection{Experiment Setup}

\textbf{Dataset and Model.} We explore two widely-used benchmark datasets in FL: CIFAR-10 \cite{krizhevsky2009learning} and Shakespeare. We evaluated the performance of SimpleCNN \cite{mcmahan2017communication} and VGG11 \cite{simonyan2014very} models on the CIFAR-10 dataset for image classification tasks, as well as the performance of an LSTM \cite{kim2016character} model on the Shakespeare dataset for character prediction tasks. 

\textbf{Baselines and Metric.} FedFQ can be integrated as an additional module into any FL framework. In this work, we choose the classic FedAvg algorithm in federated learning as the baseline. We compare FedFQ with the current state-of-the-art compressed FL algorithms, including the quantization algorithms FedPAQ \cite{reisizadeh2020fedpaq} and AQG \cite{mao2022communication} and the sparsification quantization hybrid algorithm AC-SGD \cite{yan2022ac}. Our evaluation metrics include accuracy and convergence speed at a fixed compression ratio, as well as the required communication volume to achieve the target accuracy.

\textbf{Setup of FL.} In our experiment, we uniformly set the parameters for FL. Before each communication round, the clients performed 5 local updates. The dataset is divided among 100 clients according to the rules specified in \cite{mcmahan2017communication,zhao2018federated}. To make the evaluation more convincing under the Non-IID setting, we adopted the most stringent heterogeneity configuration. Specifically, each client only receives data partitions from a single class.
For each communication round, 10 clients were randomly selected to participate in the model update process. For CIFAR-10, we set the batch size to 50 and the learning rate to 0.15. For Shakespeare, we set the batch size to 10 and the learning rate to 1.47. We evaluated the performance separately under both independent and IID and Non-IID settings in each experiment. 

\textbf{Setup of CGSA.} We set the initial temperature to 1000 and the cooling rate to 0.95. The number of iterations should be adjusted based on the parameter scale; in our experiment, it was set to 100.

Please note that due to the adoption of a high degree of heterogeneity setting and a limited number of communication rounds, our accuracy is not the current SOTA. However, our method is designed to reduce the accuracy degradation caused by compression in Non-IID settings and our goal is to compare the communication efficiency and convergence performance, rather than achieving the best accuracy on this particular task. This point can be referenced from \cite{mcmahan2017communication, zhao2018federated, yang2021cfedavg}.

\subsection{Ablation Experiment}

\begin{figure}[htbp]
	\centering
	
	\subcaptionbox{SimpleCNN-IID\label{tex2kidyangcan}}{
		
		\includegraphics[width=0.473\linewidth]{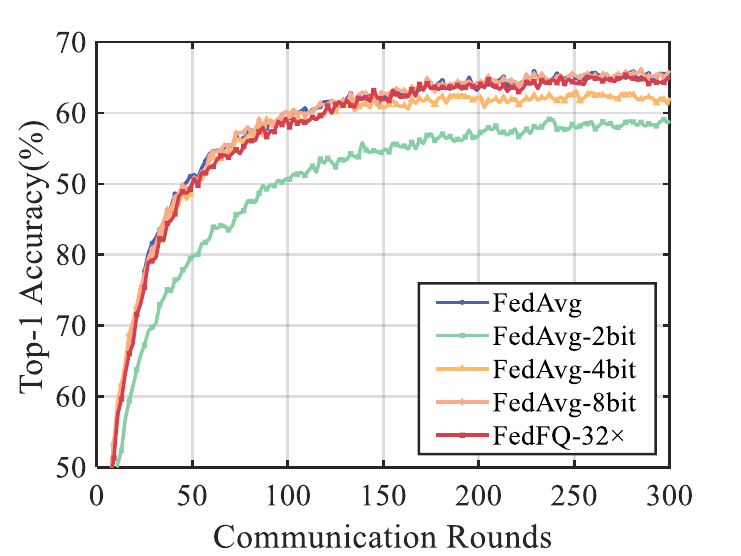}
		
	}
	\subcaptionbox{SimpleCNN-Non-IID\label{tex2kidyangcan}}{
		\includegraphics[width=0.473\linewidth]{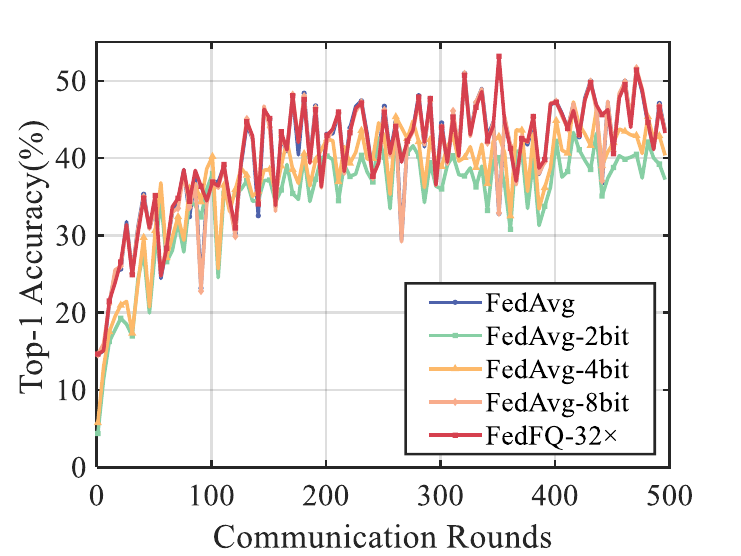}	
	}
	\subcaptionbox{VGG11-IID\label{tex2kidyangcan}}{
		\includegraphics[width=0.473\linewidth]{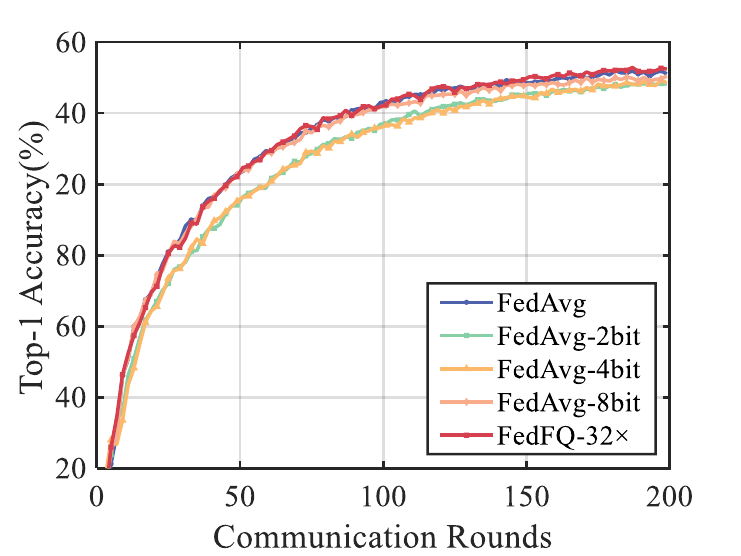}	
	}
	\subcaptionbox{VGG11-Non-IID\label{tex2kidyangcan}}{
		\includegraphics[width=0.473\linewidth]{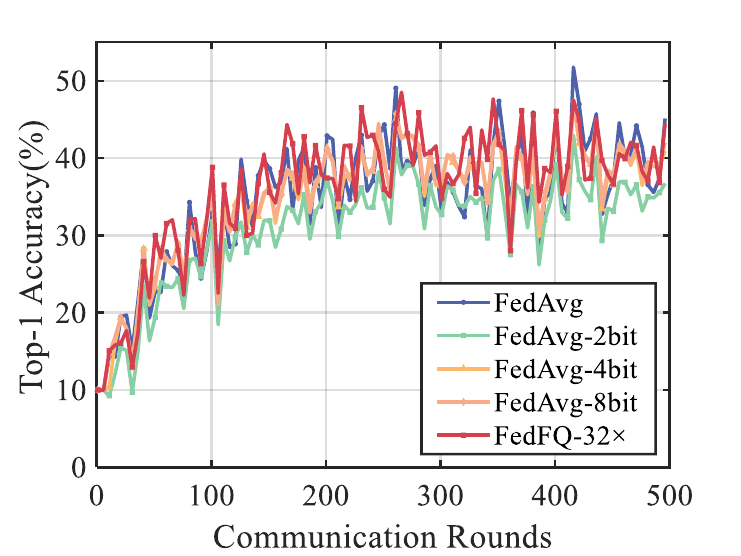}	
	}
	
	\caption{The learning curves of FedFQ and single-precision quantization. }
	
	\label{p1}
	\vspace{-15pt}
\end{figure}

We first evaluated the performance improvement of FedFQ compared to single-precision quantization algorithm. We conducted ablation experiments on three tasks under both IID and Non-IID settings. Table \ref{t1} provides the Top-1 accuracy of the global model at a fixed compression ratio, and Figure \ref{p1} illustrates the convergence curves of the training process.
\begin{table*}[htpb]
	
	\normalsize
	
	\centering

	\renewcommand{\arraystretch}{1.1}
	\setlength{\tabcolsep}{2mm}{
		\begin{tabular}{c|cccc|cccc|cc}
			
			\hline
			\multirow{2}{*}{Task} & \multicolumn{4}{c|}{SimpleCNN on CIFAR-10}                                                         & \multicolumn{4}{c|}{VGG11 on CIFAR-10}                                                             & \multicolumn{2}{c}{LSTM on Shakespeare} \\ \cline{2-11} 
			& \multicolumn{2}{c|}{IID}                                   & \multicolumn{2}{c|}{Non-IID}          & \multicolumn{2}{c|}{IID}                                   & \multicolumn{2}{c|}{Non-IID}          & \multicolumn{2}{c}{Non-IID}             \\ \hline
			Method                & \multicolumn{1}{c|}{Acc.(\%)} & \multicolumn{1}{c|}{Comp.} & \multicolumn{1}{c|}{Acc.(\%)} & Comp. & \multicolumn{1}{c|}{Acc.(\%)} & \multicolumn{1}{c|}{Comp.} & \multicolumn{1}{c|}{Acc.(\%)} & Comp. & \multicolumn{1}{c|}{Acc.(\%)}  & Comp.  \\ \hline
			FedAvg                & \multicolumn{1}{c|}{77.78}    & \multicolumn{1}{c|}{1$\times$}    & \multicolumn{1}{c|}{51.19}    & 1$\times$    & \multicolumn{1}{c|}{75.77}    & \multicolumn{1}{c|}{1$\times$}    & \multicolumn{1}{c|}{45.74}    & 1$\times$    & \multicolumn{1}{c|}{47.85}     & 1$\times$     \\ \hline
			FedAvg-2bit           & \multicolumn{1}{c|}{74.54}    & \multicolumn{1}{c|}{16$\times$}   & \multicolumn{1}{c|}{44.28}    & 16$\times$   & \multicolumn{1}{c|}{74.26}    & \multicolumn{1}{c|}{16$\times$}   & \multicolumn{1}{c|}{42.97}    & 16$\times$   & \multicolumn{1}{c|}{44.51}     & 16$\times$    \\ \hline
			FedAvg-4bit           & \multicolumn{1}{c|}{76.4}     & \multicolumn{1}{c|}{8$\times$}    & \multicolumn{1}{c|}{47.36}    & 8$\times$    & \multicolumn{1}{c|}{74.62}    & \multicolumn{1}{c|}{8$\times$}    & \multicolumn{1}{c|}{44.58}    & 8$\times$    & \multicolumn{1}{c|}{46.29}     & 8$\times$     \\ \hline
			FedAvg-8bit           & \multicolumn{1}{c|}{77.37}    & \multicolumn{1}{c|}{4$\times$}    & \multicolumn{1}{c|}{51.12}    & 4$\times$    & \multicolumn{1}{c|}{75.54}    & \multicolumn{1}{c|}{4$\times$}    & \multicolumn{1}{c|}{45.39}    & 4$\times$    & \multicolumn{1}{c|}{47.73}     & 4$\times$     \\ \hline
			\textbf{FedFQ-32$\times$}             & \multicolumn{1}{c|}{77.68}    & \multicolumn{1}{c|}{32$\times$}   & \multicolumn{1}{c|}{51.43}    & 32$\times$   & \multicolumn{1}{c|}{76.18}    & \multicolumn{1}{c|}{32$\times$}   & \multicolumn{1}{c|}{46.11}    & 32$\times$   & \multicolumn{1}{c|}{48.12}     & 32$\times$    \\ \hline
			\textbf{FedFQ-64$\times$}             & \multicolumn{1}{c|}{76.62}    & \multicolumn{1}{c|}{64$\times$}   & \multicolumn{1}{c|}{48.54}    & 64$\times$   & \multicolumn{1}{c|}{74.13}    & \multicolumn{1}{c|}{64$\times$}   & \multicolumn{1}{c|}{44.38}    & 64$\times$   & \multicolumn{1}{c|}{46.11}     & 64$\times$    \\ \hline
			\textbf{FedFQ-128$\times$}            & \multicolumn{1}{c|}{74.41}    & \multicolumn{1}{c|}{128$\times$}  & \multicolumn{1}{c|}{45.30}     & 128$\times$  & \multicolumn{1}{c|}{72.77}    & \multicolumn{1}{c|}{128$\times$}  & \multicolumn{1}{c|}{42.21}    & 128$\times$  & \multicolumn{1}{c|}{44.90}      & 128$\times$   \\ \hline
		\end{tabular}
		\caption{The Top-1 accuracy comparison between FedFQ and single-precision quantization.}
		\label{t1}
	}
\end{table*}
It can be observed that lower quantization bit-width lead to a significant decrease in model accuracy and convergence performance. 
FedFQ achieves lossless accuracy and convergence performance at nearly 32$\times$ higher compression ratios compared to single-precision quantization. Compared to 2-bit quantization, which possesses the highest compression ratio, FedFQ not only significantly surpasses in convergence performance but also improves communication compression ratios by nearly double. In addition, we also observed that as the compression ratio increases, the convergence performance decreases, requiring the model to undergo more iterations. Therefore, computation and communication are often a trade-off. FedFQ can better balance communication and performance, thanks to its fine-grained quantization.

\subsection{Comparative Experiment}
\subsubsection{The Experiments On CIFAR-10}

The CIFAR-10 dataset consists of 50,000 training examples. We employed the dataset partitioning strategy described in \cite{mcmahan2017communication,zhao2018federated}. For the IID case, an equal number of samples are randomly assigned to each client. For the Non-IID case, we sorted the dataset based on the numerical labels and divided it into data shards, which were then assigned to the clients.
\begin{figure}[htbp]
	\centering
	
	\subcaptionbox{IID\label{tex2kidyangcan}}{
		
		\includegraphics[width=0.473\linewidth]{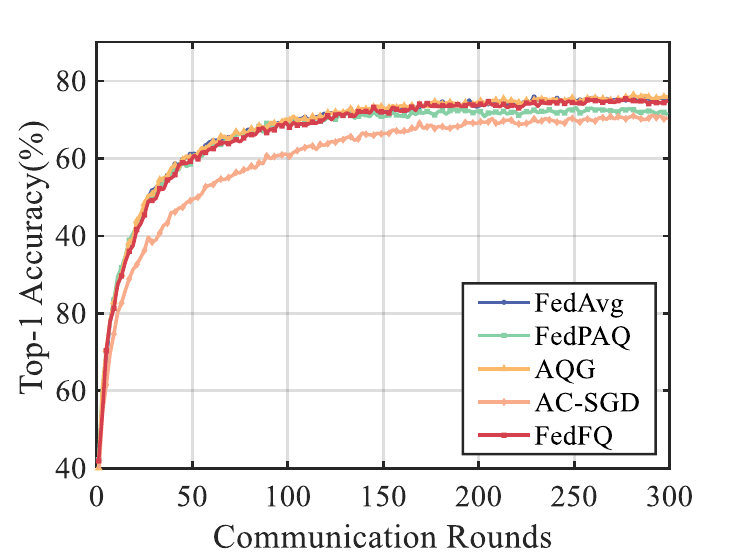}
		
	}
	\subcaptionbox{IID\label{tex2kidyangcan}}{
		\includegraphics[width=0.473\linewidth]{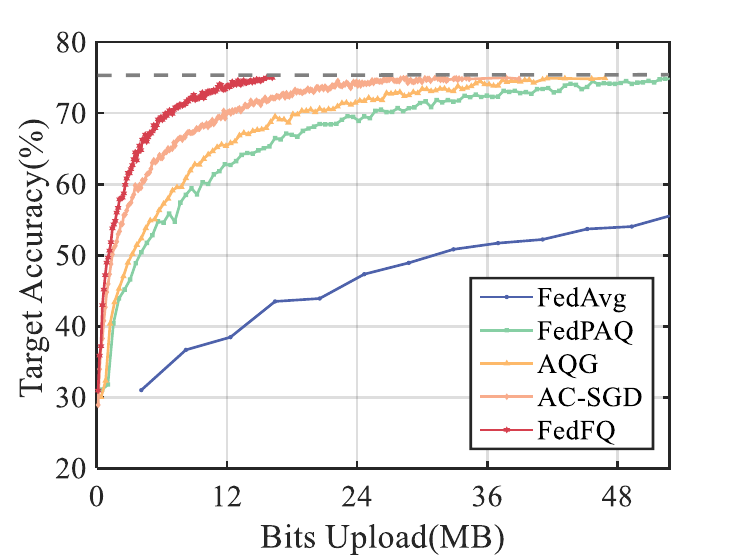}	
	}
	\subcaptionbox{Non-IID\label{tex2kidyangcan}}{
		\includegraphics[width=0.473\linewidth]{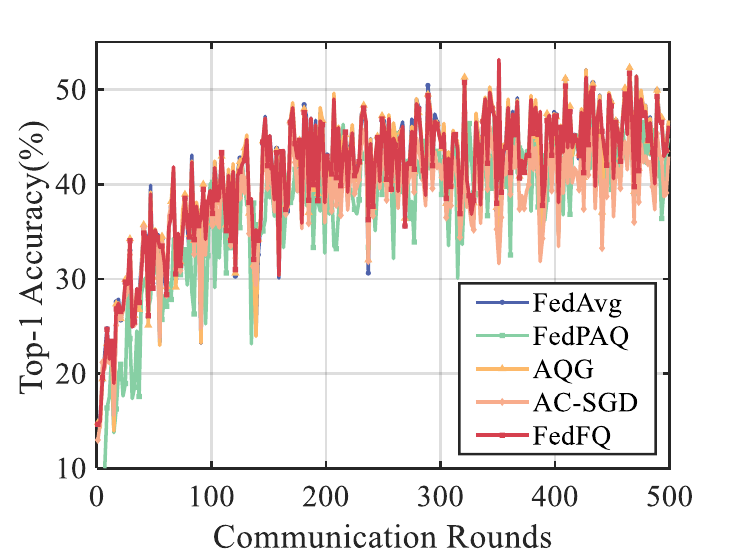}	
	}
	\subcaptionbox{Non-IID\label{tex2kidyangcan}}{
		\includegraphics[width=0.473\linewidth]{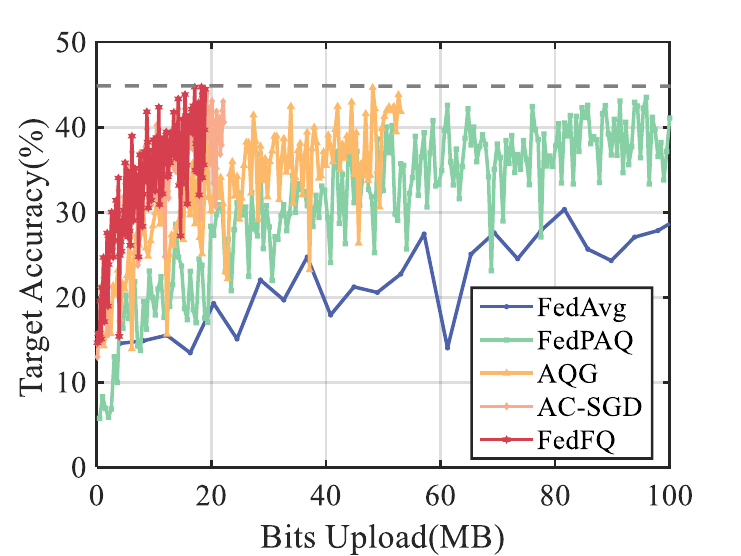}	
	}
	
	\caption{Performance comparison between FedFQ and comparison algorithms on SimpleCNN.}
	
	\label{p2}
\end{figure}
\begin{figure}[htbp]
	\centering
	
	\subcaptionbox{IID\label{tex2kidyangcan}}{
		
		\includegraphics[width=0.473\linewidth]{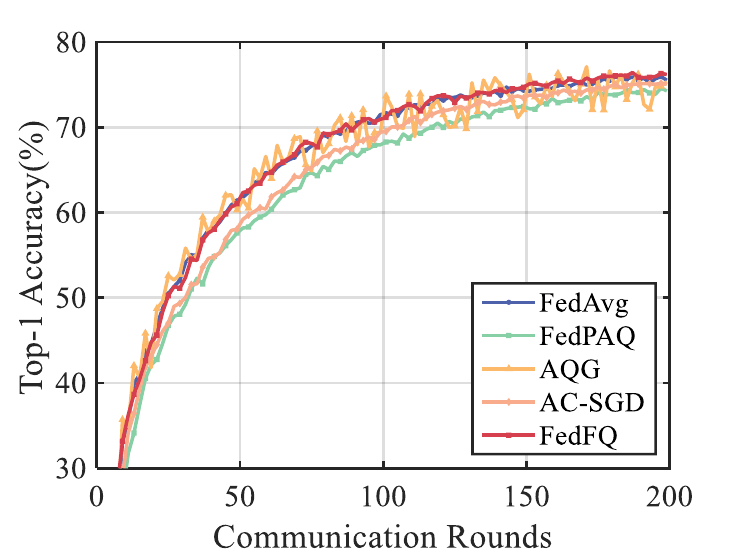}
		
	}
	\subcaptionbox{IID\label{tex2kidyangcan}}{
		\includegraphics[width=0.473\linewidth]{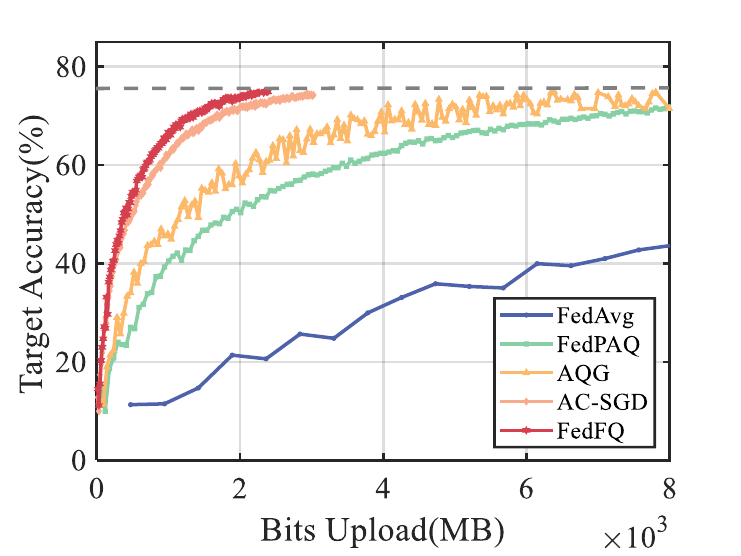}	
	}
	\subcaptionbox{Non-IID\label{tex2kidyangcan}}{
		\includegraphics[width=0.473\linewidth]{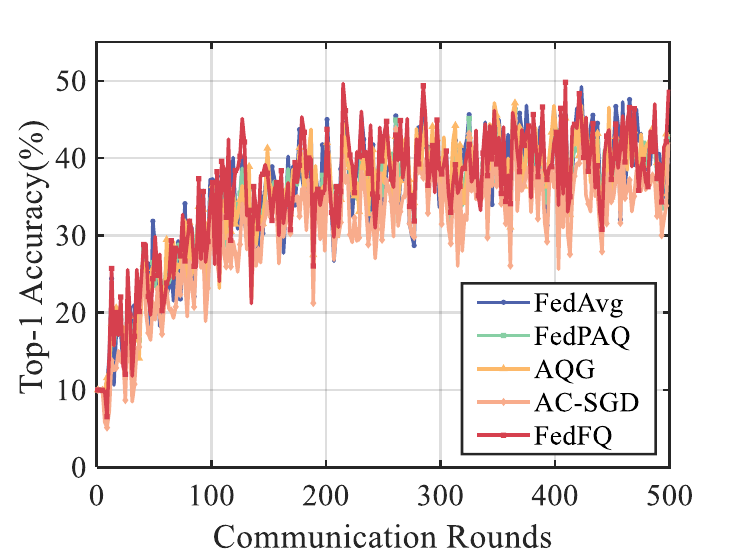}	
	}
	\subcaptionbox{Non-IID\label{tex2kidyangcan}}{
		\includegraphics[width=0.473\linewidth]{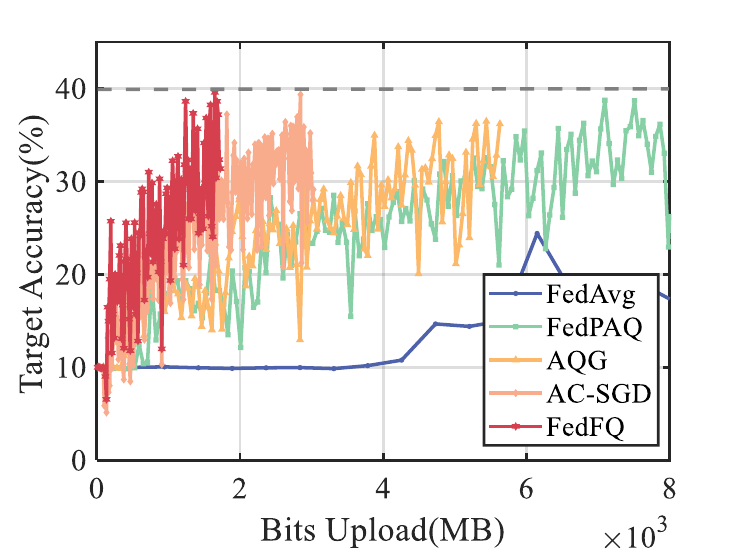}	
	}
	
	\caption{Performance comparison between FedFQ and comparison algorithms on VGG11.}
	
	\label{p3}
\end{figure}
\begin{table*}[htpb]
	
	\normalsize
	\centering
	\renewcommand{\arraystretch}{1.1}
	\setlength{\tabcolsep}{2mm}{
		\begin{tabular}{c|cc|cc|c}
			\hline
			Task        & \multicolumn{2}{c|}{SimpleCNN on CIFAR-10} & \multicolumn{2}{c|}{VGG11 on CIFAR-10} & LSTM on Shakespeare \\ \hline
			Target Acc. & \multicolumn{1}{c|}{75\%}      & 45\%      & \multicolumn{1}{c|}{75\%}    & 40\%    & 50\%                \\ \hline
			Method      & \multicolumn{1}{c|}{IID}       & Non-IID   & \multicolumn{1}{c|}{IID}     & Non-IID & Non-IID             \\ \hline
			FedAvg      & \multicolumn{1}{c|}{416.34MB}  & 571.44MB  & \multicolumn{1}{c|}{63.75GB} & 58.67GB & 53.18MB             \\ \hline
			FedPAQ      & \multicolumn{1}{c|}{52.55MB}   & 95.92MB   & \multicolumn{1}{c|}{10.51GB} & 10.05GB & 10.95MB             \\ \hline
			AQG         & \multicolumn{1}{c|}{40.41MB}   & 48.16MB   & \multicolumn{1}{c|}{6.24GB}  & 6.88GB  & 4.38MB              \\ \hline
			AC-SGD      & \multicolumn{1}{c|}{27.04MB}   & 22.07MB   & \multicolumn{1}{c|}{2.96GB}  & 2.78GB  & 2.60MB              \\ \hline
			\textbf{FedFQ}       & \multicolumn{1}{c|}{15.15MB}   & 18.93MB   & \multicolumn{1}{c|}{2.33GB}  & 1.61GB  & 0.90MB              \\ \hline
		\end{tabular}
		\caption{The number of bits required to achieve the target accuracy.}
		\label{t2}}
	
\end{table*}

We evaluated the performance of FedFQ on both SimpleCNN and VGG11. SimpleCNN has a simple structure with a smaller number of parameters, while VGG11 has a larger parameter count and is widely used in real-world tasks. We evaluated the performance based on two metrics: (1) Convergence performance and (2) Communication cost required to achieve the target accuracy. Figure \ref{p2} and Figure \ref{p3} present the performance of FedFQ on these two networks, respectively. It can be observed that  FedFQ exhibits optimal convergence performance and achieves the minimum communication required to reach the target accuracy. Under the IID setting, FedFQ achieves a communication compression of nearly 27$\times$ compared to the baseline experiment. Under the Non-IID setting, FedFQ achieves a communication compression of nearly 36$\times$ compared to the baseline experiment. The results can be found in Table \ref{t2}.

\subsubsection{The Experiments On Shakespeare}

We performed a character prediction task on the Shakespeare dataset using a stacked character-level LSTM model. This model processes sequences of characters by embedding each character into an 8-dimensional space and then passing them through two LSTM layers, each containing 256 nodes.

\begin{figure}[htbp]
	\centering
	
	\subcaptionbox{Non-IID\label{tex2kidyangcan}}{
		
		\includegraphics[width=0.473\linewidth]{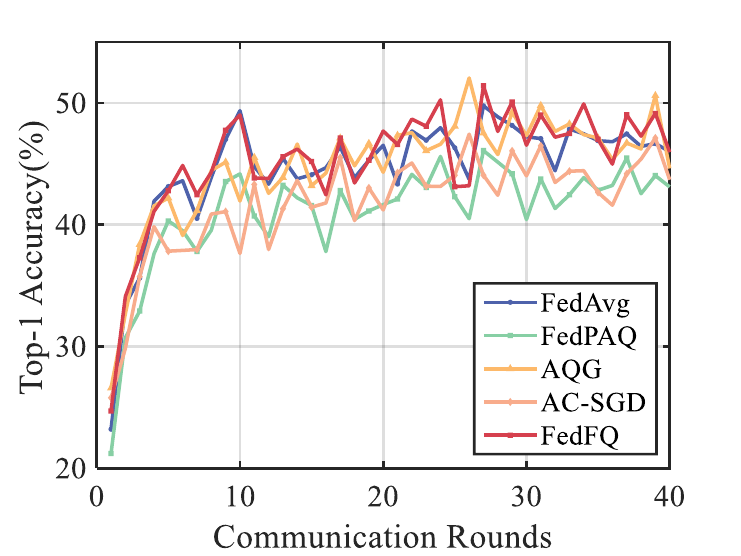}
		
	}
	\subcaptionbox{Non-IID\label{tex2kidyangcan}}{
		\includegraphics[width=0.473\linewidth]{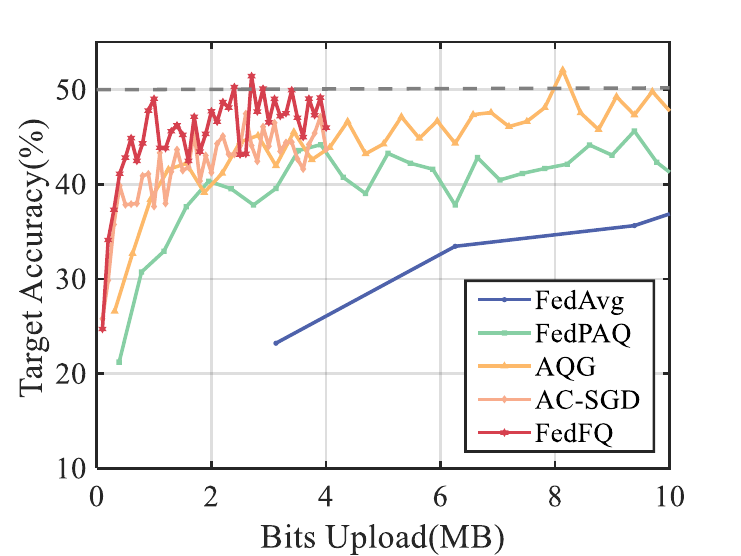}	
	}

	\caption{Performance comparison between FedFQ and comparison algorithms on LSTM.}
	\vspace{-10pt}
	\label{p4}
\end{figure}

In the Non-IID setting, we followed a predefined data splitting scheme detailed in \cite{mcmahan2017communication}.
Table \ref{t2} presents the communication overhead required to achieve the target accuracy. FedFQ achieves a communication compression of nearly 63$\times$ compared to the baseline experiment. This performance surpasses the minimal communication overhead hybrid compression method. The convergence curves of FedFQ and the comparison methods are shown in Figure \ref{p4}.

\subsection{Evaluation in the Real-world Network Environment}

To enhance the applicability and relevance of FedFQ, we evaluated FedFQ in a real network using NVIDIA RTX3090 GPUs as clients in a FL setup. We conducted image classification tasks on the CIFAR-10 dataset with a ResNet-20 model and a batch size of 64. Data was partitioned according to the method in \cite{mcmahan2017communication}. Aggregation occurred every fifth local update. We measured the average training time per epoch and network speed for varying client numbers. Results without FedFQ are in Table \ref{t3}, and those with FedFQ are in Table \ref{t4}.
\begin{table}[htpb]
	
	\normalsize
	\centering
	\renewcommand{\arraystretch}{1.1}
	\setlength{\tabcolsep}{2mm}{
		\begin{tabular}{c|c|cll}
			\cline{1-3}
			Internet   Speed & Clients & Training Time &  &  \\ \cline{1-3}
			32.87Mbps         & 2       & 45.1s         &  &  \\ \cline{1-3}
			30.43Mbps         & 4       & 28.4s         &  &  \\ \cline{1-3}
			35.33Mbps         & 8       & 22.3s         &  &  \\ \cline{1-3}
			32.18Mbps         & 16      & 18.6s        &  &  \\ \cline{1-3}
		\end{tabular}
		\caption{The training time for each epoch without deploying FedFQ.}

		\label{t3}}
	\vspace{-5pt}
\end{table}
\begin{table}[htpb]
	
	\normalsize
	\centering
	\renewcommand{\arraystretch}{1.1}
	\setlength{\tabcolsep}{2mm}{
		\begin{tabular}{c|c|cll}
			\cline{1-3}
			Internet   Speed & Clients & Training Time &  &  \\ \cline{1-3}
			35.74Mbps         & 2       & 47.8s         &  &  \\ \cline{1-3}
			33.15Mbps         & 4       & 26.8s         &  &  \\ \cline{1-3}
			33.22Mbps         & 8       & 14.6s         &  &  \\ \cline{1-3}
			30.96Mbps         & 16      & 9.42s        &  &  \\ \cline{1-3}
		\end{tabular}
		\caption{The training time for each epoch after deploying FedFQ.}
		
		\label{t4}}
	\vspace{-5pt}
\end{table}

Through our experiments, we observed that FedFQ did not achieve acceleration when the number of clients was small. This was because, at that point, the majority of the training time was dominated by computation, and the reduction in communication time provided by FedFQ was negligible in comparison. As the number of clients increased, the computational power significantly improved. However, the simultaneous transmission of parameter updates from multiple clients to the server led to network congestion. At this stage, communication time became the dominant factor. FedFQ substantially reduced the communication overhead, thereby accelerating the training process. Therefore, FedFQ is of significant importance for large-scale FL in real network environments.
\section{Conclusion}
This paper has proposed FedFQ, a communication-efficient FL algorithm with a fine-grained adaptive quantization strategy to address the trade-off between high communication compression ratios and superior convergence performance. We have conducted a rigorous theoretical analysis of the convergence of FedFQ and proposed a parameter-level mixed precision quantization strategy with polynomial-level complexity based on the theory. Extensive experiments based on benchmark datasets have
been conducted to verify the effectiveness of the proposed
algorithm and demonstrate that our algorithm has minimal communication overhead without sacrificing performance.
In particular, when converging to the same accuracy, FedFQ achieves a compression of 27$\times$ to 63$\times$ in terms of upstream communication volume compared to the baseline experiment.

\bibliography{aaai25}

\end{document}